\documentclass{aa}
\usepackage{graphicx}

\begin{document}

\title{Superoutbursts, superhumps and the tidal-thermal instability model}
\titlerunning{Superoutbursts, superhumps and the TTI Model}
\authorrunning{Buat-M\'enard \and Hameury}
\author{Valentin Buat-M\'enard \and Jean-Marie Hameury}
\offprints{V. Buat-M\'enard}
\mail{buat@astro.u-strasbg.fr}

\institute{UMR 7550 du CNRS, Observatoire de Strasbourg, 11 rue de
l'Universit\'e, F-67000 Strasbourg, France}
\date{Received / Accepted }

\abstract{We include the tidal instability due to the 3:1
resonance in the disc instability model developed by Hameury et
al. (\cite{hmd98}) and modified by Buat-M\'enard et al.
(\cite{b01}). We confirm earlier results by Osaki (\cite{o89})
that the tidal instability can account for the SU UMa light
curves. We show that in ultra-low mass ratio systems such as ER
UMa stars and WZ Sge stars, the superoutburst ends while the disc
is still eccentric, as proposed by Hellier (\cite{h01}). However,
since the disc shrinks rapidly once a cooling wave has started,
the eccentricity should stop shortly after the end of a
superoutburst. This result disagrees with the suggestion by
Hellier that decoupling the thermal and tidal instability in the
TTI model can account for late superhumps and echo outbursts in
ultra-low mass ratio systems. We propose instead that ER UMa short
supercycles can be explained either by the alternation of narrow
and wide outbursts similar to those occurring in \object{SS Cyg},
or by the effects of irradiation (Hameury et al. \cite{hlw00}). In
both cases, we predict that superhumps should be permanent, which
is suggested by observations (Gao et al. \cite{g99}). We can also
reproduce light curves similar to those of \object{EG Cnc},
varying the mass transfer rate in a TTI model including both
irradiation and the presence of an inner hole in the disc.
 }
\maketitle

\section{Introduction}

Dwarf novae are cataclysmic variables (CVs) which exhibit 2-6 mag
recurrent outbursts (see Warner \cite{w95} for an encyclopaedic
review). It is generally accepted that these outbursts are due to
a thermal-viscous instability that occurs in an accretion disc in
which the viscosity is given by the $\alpha$-prescription (Shakura
\& Sunyaev \cite{s73}). The instability arises when hydrogen
becomes partially ionized and the opacities vary steeply with
temperature; this occurs when the disc material reaches
temperatures of order of 8000 K (see Lasota \cite{l01} for a
review of the model). A number of characteristics of many light
curves (characteristic time scales, amplitudes, ...) are
reproduced by the disc instability model (DIM) provided the
viscosity parameter $\alpha$ is different in quiescence and in
outburst. However several subclasses of dwarf novae show specific
features that are not easy to explain with the standard version of
the DIM.

SU UMa stars are defined as dwarf novae in which normal outbursts
cycles are regularly interrupted by anomalously bright and long
outbursts called superoutbursts. These outbursts are ~0.7
magnitudes brighter and last 5 to 10 times longer than the normal
ones. The mean time $T_s$ between two consecutive superoutbursts
defines the period of a supercycle $T_s$; in most SU UMa stars,
$130 < T_s <350$ days. There are however two extreme exceptions:
ER UMa stars have anomalously short supercycles (between 19 and 45
days), while WZ Sge stars show only superoutbursts separated by
very long quiescence periods that can reach 30 years. In addition,
the light curve is modulated during superoutbursts at a period
slightly longer than the orbital period. These so-called
superhumps usually disappear shortly after the decline of the
superoutburst. Whitehurst (\cite{w88}) explained these superhumps
by the precession of a distorted disc. The name `superoutburst' is
reserved for dwarf nova eruptions in which a superhump is present.
This property differentiates superoutbursts from wide outbursts
observed in many dwarf novae; van Paradijs (\cite{vanP83})
suggested that superoutbursts are just wider outbursts in the
bimodal outburst-length distribution. In addition, a 1985 outburst
of U Gem had all the properties of a superoutburst (of WZ Sge
stars) except for the superhump (see Kuulkers et al. \cite{ku99}
and references therein). This could suggest that the superoutburst
phenomenon is independent of the disc distortion. We discuss this
possibility in the Section 4.

As almost all SU UMa stars have periods below the period gap of
CVs, they all have low secondary to primary mass ratio $q$, and
the primary Roche lobe is large.  The disc can therefore extend to
relatively large radii, and this led Osaki (\cite{o89}) to propose
the thermal-tidal instability (TTI), in which superoutbursts are
caused by the disc reaching the 3:1 resonance radius, as a model
for SU UMa stars. SPH simulations by Whitehurst (\cite{w88})
showed that the disc becomes eccentric and precesses when its
radius reaches the 3:1 resonance radius, but it must be kept in
mind that the hydrodynamical simulations of Stehle (\cite{s99})
and Kornet \& R\'o\.zyczka (\cite{kr00}) did not show such an
effect.

Detailed observations of SU UMa stars point out several
limitations of the TTI model. The mass transfer rate from the
secondary is observed to be enhanced during outburst; this
increase, possibly due to irradiation, is not taken into account,
although it can drastically alter the light curves (Hameury et al.
\cite{hlw00}).

The TTI model cannot easily reproduce the properties of ER UMa and
WZ Sge subclasses. In order to obtain the very short ER UMa
supercycles (19 to 45 days), one must assume that the tidal
instability shuts off prematurely (Osaki \cite{o95}) for very
small mass ratios. The very long quiescence time of WZ Sge systems
would on the other hand require a very small viscosity in
quiescence (Smak \cite{s93}).

Occasionally, some systems show echo outbursts at the end of a
superoutburst: several consecutive small outbursts with very short
recurrence time are triggered before the disc returns to
quiescence.  During these echo-outbursts, superhumps are still
present. These so-called late superhumps (LS) have also been
observed in ER UMa stars. However, the TTI model simulations
predict that the end of a superoutburst should coincide with the
cessation of the tidal instability.

These different problems have lead several authors to investigate
alternative possibilities. Hameury et al. (\cite{hlw00}) have
shown that if one takes into account the disc and secondary
illumination in the standard DIM, one is able to reproduce light
curves reminiscent of those of systems such as \object{RZ LMi} or
\object{EG Cnc}. Recently Hellier (\cite{h01}) suggested that some
difficulties could be solved in the framework of a slightly
modified TTI model; he proposed that the superoutburst must end
before the end of eccentricity in low mass ratio systems, thus
giving an explanation for LS.

We have included in our DIM model (Hameury et al. \cite{hmd98};
Buat-M\'enard et al. \cite{b01}, hereafter Paper I) an enhanced
tidal torque when the disc radius exceeds the 3:1 resonance
radius. We do reproduce the earlier results of Osaki (\cite{o89}),
obtained using a simplified model. We show that in a number of
cases, a cooling wave can end a superoutburst while the disc is
still eccentric, thereby confirming the proposal of Hellier
(\cite{h01}). However, this by itself is not sufficient to
reproduce the properties of ER UMa or WZ Sge systems, and
additional effects such as illumination and mass transfer
variations must be included.

\section{The TTI model}

The equation of angular momentum conservation in a disc is
(Hameury et al. \cite{hmd98}):
\begin{eqnarray}
j \frac{\partial \Sigma}{\partial t} = - \frac{1}{r} \frac{\partial}{\partial r}
\left(r \Sigma j v_{\rm r}\right) & + & \frac{1}{r} \frac{\partial}{\partial r}
\left(- \frac{3}{2} r^2 \Sigma \nu \Omega_{\rm K} \right) \nonumber\\
& + & \frac{j_2}{2 \pi r}\frac{\partial \dot{M}_{\rm tr}}{\partial
r} - \frac{1}{2 \pi r} T_{\rm tid}(r) \label{eq1}
\end{eqnarray}
where $\Sigma$ is the surface column density, $\dot{M}_{\rm tr}$
is the rate at which mass is incorporated into the disc at radius
$r$, $v_{\rm r}$ is the radial velocity in the disc, $j = (G M_1
r)^{1/2}$ is the specific angular momentum of material at radius
$r$ in the disc, $\Omega_{\rm K} = (G M_1/ r^3)^{1/2}$ is the
Keplerian angular velocity (where $M_1$ is the primary mass),
$\nu$ is the kinematic viscosity coefficient, and $j_2$ the
specific angular momentum of the material transferred from the
secondary. $T_{\rm tid}$ is the torque due to the tidal forces,
for which we use the formula of Smak (\cite{s84}), derived from
the determination of tidal torques by Papaloizou \& Pringle
(\cite{pp77}):
\begin{eqnarray}
T_{\rm tid} = c \omega r \nu \Sigma \left(\frac{r}{a}\right)^5 \label{eq2}
\end{eqnarray}
where $\omega$ is the angular velocity of the binary orbital motion, $a$ the
binary orbital separation and $c$ a numerical constant taken so as to give an
average disc radius equal to some chosen value.

\begin{figure}
\resizebox{\hsize}{!}{\includegraphics[angle=-90]{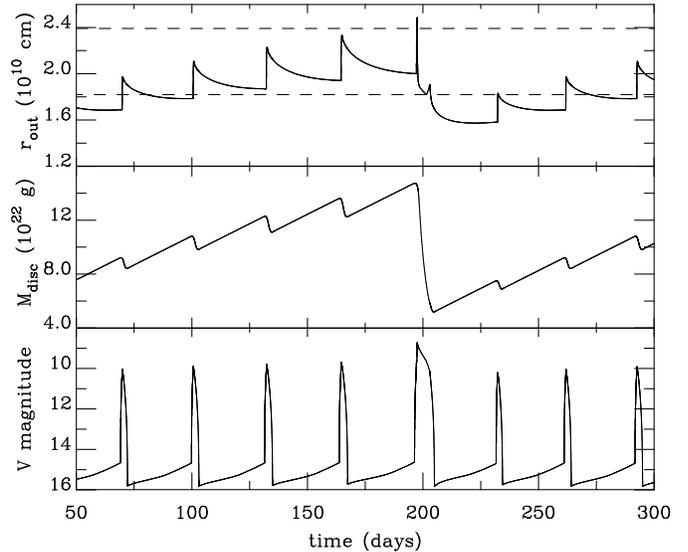}}
\caption{Superoutburst cycle in the TTI model for SU UMa with
$\dot{M}_{\rm tr} = 10^{16}$g.s$^{-1}$. Top panel: disc radius;
intermediate panel: disc mass; bottom panel: disc visual
magnitude} \label{fig1}
\end{figure}

In the TTI model, $c$ is no longer a constant, but depends on
whether the disc is eccentric or not. Ichikawa et al.
(\cite{iho93}) reproduced supercycles assuming that the disc
becomes eccentric when its radius reaches $r_{3:1} \sim 0.46 a$;
$c$ is then increased by a large factor. Afterward the disc starts
shrinking and can no longer maintain its eccentricity when its
radius becomes smaller than $r_{\rm crit0} \sim 0.35 a$, at which
point $c$ returns to its normal value. We choose to take a value
$c_0$ when the disc is axisymmetric and a larger value $c_1$ when
the disc is eccentric. This is our Prescription $I$ for $c$, quite
similar to that of Ichikawa et al. (\cite{iho93}) who extended the
work of Osaki (\cite{o89}) by taking into account a non zero tidal
torque during quiescence.

We have included this variable tidal torque in the disc
instability model used in paper I, which is a modified version of
the DIM described in Hameury et al. (\cite{hmd98}) that includes
heating by the stream impact and the tidal torque (We have
corrected Eq. (1) of paper I, in which the heating rate was
overestimated by a factor $\sim$ 2). In the following, we also
assume that $\alpha = \alpha_{\rm cold} = 0.04$ in quiescence and
$\alpha = \alpha_{\rm hot} = 0.2$ in outburst. The inner radius is
fixed at $r_{\rm in} = 10^9$ cm.

We obtain the cycle shown in Fig. \ref{fig1}. After a
superoutburst, the disc mass and radius are small. The disc mass
slowly increases, and several normal outbursts are triggered due
to the thermal instability. During each outburst, only a small
amount of mass is accreted by the white dwarf, so that the average
disc mass increases; a significant disc expansion occurs during
the rise to maximum, when the heat front reaches the outer edge of
the disc. The disk then shrinks when a cooling front brings the
disc back to a cool state, causing a significant reduction of the
outward angular momentum flow. After the end of an outburst, the
disc radius very slowly decreases as a result of mass addition and
viscous diffusion, but the net balance is an increase of the
radius from one outburst to the next one. Eventually, $r_{\rm
out}$ becomes larger than $r_{3:1}$ during an outburst and the
tidal torque is increased, as well as the associated energy
dissipation. The disc is then maintained in the hot state until
most of the mass has been accreted. During this phase the radius
decreases until $r < r_{\rm crit0}$. The disc returns to a
circular shape, the anomalous tidal dissipation stops, and a
cooling wave starts from the disc outer edge, bringing the whole
disc in a cool state.

\subsection{Radial dependence of $c$}

There is, however, no compelling reason to assume that the whole
disc is affected by the development of the tidal instability. We
therefore use another prescription for $c$ (hereafter Prescription
$II$). We assume that when the disc is circular, $c = c_0$; when
eccentric, $c(r) = c_1$ for $r > 0.9 \; r_{\rm crit0}$ and $c =
c_0$ for $r < 0.8 \; r_{\rm crit0}$.  A linear interpolation is
assumed between $0.8 \; r_{\rm crit0}$ and $0.9 \; r_{\rm crit0}$.
We compare and discuss the results obtained using both
prescriptions in \ref{influence}.

\subsection{Time delays}

Temporal variations of $c(r)$ are not instantaneous. Observations
show that normal superhumps appear within a few of days after the
onset of a superoutburst, and that they develop within a day, so
that it does not take long for the disc to become eccentric
(Semeniuk \cite{s80}). However, it takes the disc at least a
dynamical time to change its geometrical shape; we assumed that,
when the tidal instability sets in, $c$ increases linearly on a
timescale $t_{\rm r} = 2 \; 10^4$ s. Note that in WZ Sge systems,
the early superhumps that first appear seem to be of a different
nature from the normal superhumps; Osaki \& Meyer (\cite{om02})
suggested that they could be related to the 2:1 resonance.

Similarly, the disc must dissipate its excess energy in order to return to a
circular shape.  The relative energy difference between a circular and an
elliptic orbit with eccentricity $e$ with the same angular momentum is
$\Delta E/ E_{\rm circ} = e^2$. As the dissipation at radius $r$ is $D \sim G
M \Sigma \nu / r^3$, the time needed for the disc to return to a normal state
is:
\begin{equation}
t_{\rm e-c} = \frac{\Delta E}{D} = e^2 \frac{r^2}{2 \nu} = e^2 t_\nu
\label{eq5}
\end{equation}
where $t_\nu$ is the viscous time in the outer part of the disc.
When the disc reverts to its normal state, the cooling wave has
already started or will start with the end of tidal instability
(see section \ref{decouple}). $t_\nu$ therefore refers to the cool
state, and is of the order of the recurrence time, since the
outbursts obtained here are of the inside-out type. Typical values
are tens of days.  Numerical simulations of the development of the
tidal instability show that $e$ is of order of 0.1--0.3 (Murray
\cite{m96}), which is confirmed by observations (Patterson et al.
\cite{p00}), and $t_{\rm e-c}$ should be a few percent of $t_\nu$.
We then assume that $c$ linearly decreases when the disc returns
to a circular shape on a time $t_{\rm f} = 2 \; 10^5$ s.

We use three different sets of parameters that are representative
for \object{SU UMa}, a typical ER UMa star and a typical WZ Sge
star respectively, shown in Table \ref{tab1}. $r_{\rm circ}$ is
the circulation radius at which a particle leaving the Lagrangian
point would stay in circular orbit if there were no accretion disc
(see Paper I for details). One should note that $r_{\rm circ}$ is
calculated taking into account the gravitational potential of the
secondary, and therefore differs slightly from the Keplerian
value. In the following section, we will use the term ultra-low
mass-ratio (ULMR) stars when referring to simulations with
\object{ER UMa} parameters.

\begin{table}
\caption{\label{tab1} Parameters used for SU UMa subclasses simulations. $M_2$
is the secondary mass and $r_{\rm circ}$ the circulation radius.}
\begin{tabular}{lccc}\hline
& SU UMa & \multicolumn{2}{c}{ULMR Stars}\\
&  & ER UMa type & WZ Sge type\\
\hline
$M_1 / M_\odot$ & 0.8 & 1.0 & 0.6 \\
$M_2 / M_\odot$ & 0.15 & 0.1 & 0.1 \\
$P_{\rm orb}$ (hr) & 1.83 & 1.522 & 1.3752 \\
$a / 10^{10}$ cm & 5.2 & 4.82 & 3.876 \\
$r_{\rm circ} / 10^{10}$ cm & 0.91 & 1.1 & 0.74 \\
\hline
\end{tabular}
\end{table}

\section{Decoupling the thermal and tidal instabilities \label{decouple}}

\subsection{\label{influence} A property of ultra-low mass ratio systems ?}

Hellier (\cite{h01}) suggested that the ultra-low mass ratio of ER
UMa and WZ Sge stars is responsible for their peculiarities. In
their case, the Roche lobe extends far beyond the 3:1 resonance
radius, and the disc could in principle be larger than $r_{3:1}$.
Hellier proposed that regions outside the resonance radius suffer
a lower tidal dissipation than the portions in the 3:1 resonance
zone. Then, during a superoutburst, heating by the tidal torque
will not be as efficient as in other SU UMa stars, and a cooling
wave could start while the disc is still eccentric, and will
remain so several days after the end of the superoutburst.

This would then account for late superhumps. In addition, Hellier
suggested that this model could also explain the echo outbursts in
WZ Sge stars and the short ER UMa supercycle: at the end of a
superoutburst, the disc would be still eccentric, and the tidal
dissipation sufficiently large to trigger short outbursts during
which the heating wave does not propagate to the outer edge (one
should add that such outbursts are not only short but also have
amplitudes lower than outbursts in which the heating front reaches
the outer disc's edge). ER UMa stars would then be systems always
in an eccentric state, while WZ Sge stars would finally return to
complete quiescence with circular discs. The difference between
the supercycle duration of \object{ER UMa} and WZ Sge stars would
be due to higher mass-transfer rates in ER UMa's.

\begin{figure}
\resizebox{\hsize}{!}{\includegraphics[angle=-90]{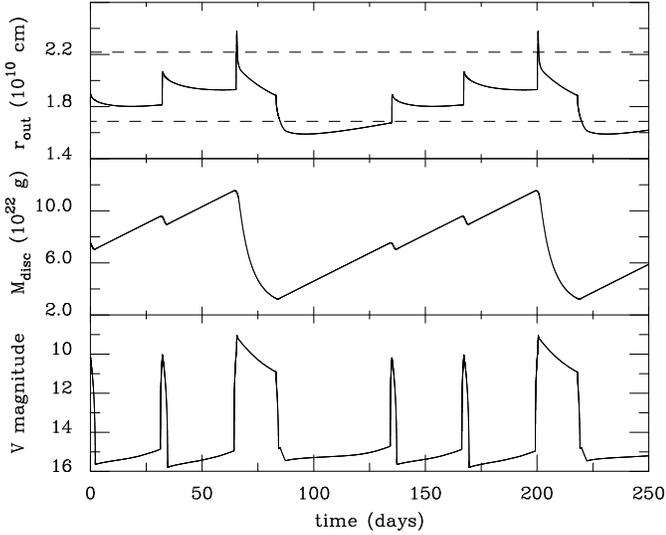}}
\caption{Same as fig 1 for ER UMa parameters.} \label{fig2}
\end{figure}

Fig \ref{fig1} shows the simulated light curve for the parameters
of SU UMa, assuming a factor 20 between $c_1$ and $c_0$. The
superoutburst ends when the disc radius becomes lower than $r_{\rm
crit0}$, i.e. the end of the tidal instability causes the end of
the superoutburst. Fig \ref{fig2} shows the simulated light curve
for the parameters of \object{ER UMa}, with the same $c_1 / c_0$.
The mass transfer rate of $10^{16}$ g s$^{-1}$ chosen here is
close to the critical value above which the system is stable
(within a factor 1.3). In this case, as predicted by Hellier, the
superoutburst ends because the thermal instability stops, while
the tidal instability is still effective. However, contrary to
Hellier's hypothesis (\cite{h01}), this phase is very short
because of the rapid shrinking of the disc. Therefore, the
duration of the supercycle is much larger than observed (there is
practically no quiescence phase in real ER UMa stars): short
outbursts are due to the usual thermal-viscous instability. As a
consequence, contrary to Hellier's assumption, the heating front
in short outbursts always reaches the outer disc's edge. Mass
transfer rates closer to the stability limit do not lead to
shorter supercycles, but rather to longer superoutbursts; this is
actually the reason for which it is impossible to obtain very
short supercycles in the standard TTI model without changing the
tidal instability condition (Osaki \cite{o95}).

\begin{figure}
\resizebox{\hsize}{!}{\includegraphics[angle=-90]{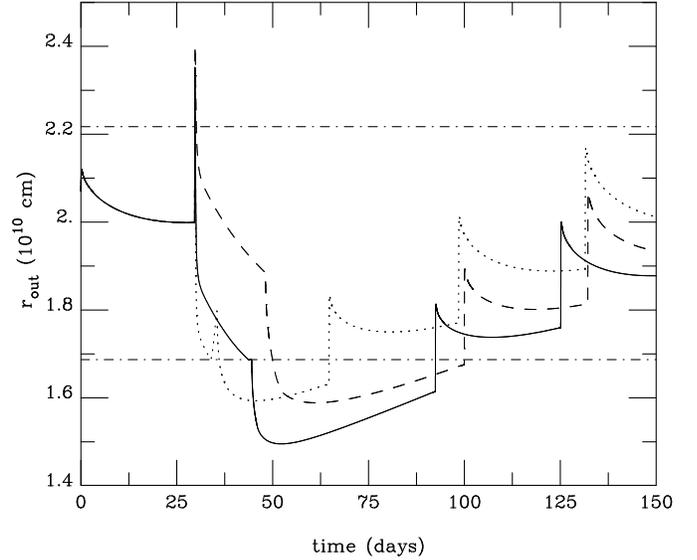}}
\caption{Radius behavior during superoutburst for ultra-low mass
ratio stars with $\dot{M}_{\rm tr} = 10^{16}$g.s$^{-1}$. The
dot-dashed lines represents the $0.46 a$ (upper) and $0.35 a$
(lower) radius. For $c_1 / c_0$ factors of 20 (dashed line) and 50
(solid line), the superoutburst lasts $\sim 21$ days ($\sim$ 17
days for $c_1 / c_0$ = 50) and is terminated by a cooling wave.
For a factor 80 (solid line) it lasts $\sim 9$ days and ends by
the disc becoming circular.} \label{fig3}
\end{figure}

We have investigated the influence of the ratio $c_1 / c_0$ and of
the Roche size on our results. Fig \ref{fig3} compares the
superoutburst behavior for $c_1 / c_0$ factors of 20, 50 and 80
for ULMR stars with prescription I. As can be seen, the width of
the superoutburst strongly depends on $c_1$. Very small and very
large values of $c_1$ both give short superoutburst, either
because the tidal heating is too small (small $c_1$), or because
the disc contraction under the effect of the tidal torque is too
large (large $c_1$). Typical $c_1/c_0$ should be in the range
$\sim$ 20 -- 50 in order to reproduce the observed superoutburst
durations.

The decoupling of the tidal and thermal instability was found for
values of $c_1/c_0$ lower than 80, for both prescription I and II
for ER UMa parameters, thereby confirming the first part of
Hellier (\cite{h01}) proposal. On the other hand, for the
parameters of \object{SU UMa}, the instability stopped because the
disc shrinks below $0.35 a$, except for $c_1/c_0< 10$; this limit
is increased to $\sim$ 20 when one uses prescription II.

These results strongly support the idea that the decoupling of
thermal and tidal instability is a property of the ultra low mass
ratio systems. One should however keep in mind that this
conclusion depends somewhat on the assumed radial dependence of
$c$, and on the assumed $c_1/c_0$; extreme values (less than 10 or
more than 50) cause too short superoutburst to be triggered but ER
UMa stars sometimes have short superoutbursts.

\subsection{The superoutburst cooling wave}

As shown in the previous section, decoupling of tidal and thermal
instabilities is expected in ultra low mass ratio SU UMa stars.
However, this is not sufficient to explain the late superhumps or
echo outbursts observed in ER UMa and WZ Sge systems, as suggested
by Hellier (\cite{h01}). When a cooling front starts to propagate,
the radius decreases rapidly (in less than a few days) to its
minimum value in the supercycle. If this minimum is less than
$r_{\rm crit0}$, then the disc will return to a circular shape in
a time scale $t_{\rm f}$, and one does not expect to observe late
superhumps. If on the other hand the minimum radius is larger than
$r_{\rm crit0}$, the disc always remains eccentric, the superhumps
are permanent, but one then loses the very mechanism that was
supposed to trigger superoutbursts.

\begin{figure}
\resizebox{\hsize}{!}{\includegraphics[angle=-90]{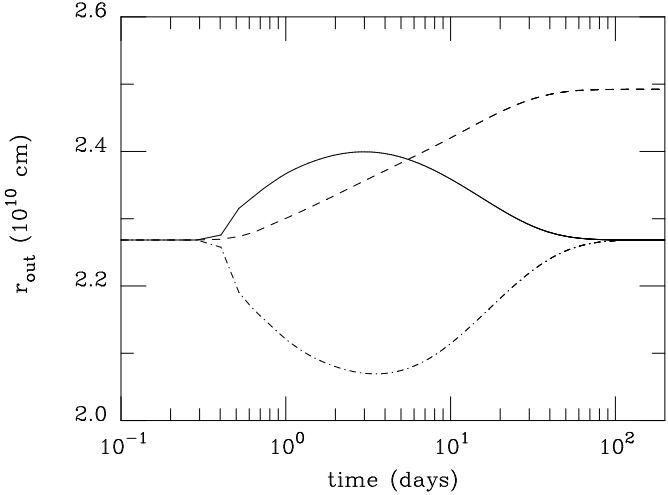}}
\caption{Responses of the outer radius of a steady accretion disc
in the hot state to variation by a factor 2 of $\dot{M}_{\rm tr}$
(solid line), $c$ (dashed line) and $\alpha$ (dot-dashed line).
The parameters vary linearly in $10^4$s. Note that the
perturbation starts at $t \sim 0.3$ d.} \label{fig4}
\end{figure}

The disc shrinkage at the end of an outburst or a superoutburst is
therefore a key question. The radius evolves under the influence
of three main effects, as shown in Eq. (\ref{eq1}): (i) the
addition of mass with low specific angular momentum at the outer
edge of the disc and (ii) the tidal torque that both act to
contract the disc, and (iii) the viscous outward transport of
angular momentum that tends to make the disc larger. These effects
are parameterized by $\dot{M}_{\rm tr}$, $c$ and $\alpha$
respectively. In order to determine their influence on the disc
radius variations, we have changed their values by a factor 2
(linear variation on a time scale of 10$^4$s) for a steady disc in
a hot state (Fig \ref{fig4}). The disc radius in the new steady
state is unchanged, except in the case where $c$ is modified; the
short term influence of these variations are summarized in table
\ref{tab2} which shows the value of $d \log r / d \log x$, $10^4$
s after the beginning of the variation, $x$ being one of the three
parameters. As can be seen, $d \log r / d \log c$ is 5 to 10 times
smaller than the two other quantities. This explains why $c_1/
c_0$ has to be larger than 10 in order to obtain a superoutburst,
and why this ratio needs to be larger than 20 in order for the
radius to decrease fast enough during a superoutburst and reach
$r_{\rm crit0}$ before the cooling wave starts.

\begin{table}
\caption{\label{tab2} Response of the outer disc radius $r_{\rm
d}$ to a decrease of $|\dot{M}|_{\rm tr}$, $c$ and $\alpha$ by a
factor 2.}
\begin{tabular}{lc}
\hline
$x$  & $d\log r_{\rm d} / d\log x$ (at $10^4$ s)\\
\hline
$|\dot{M}|_{\rm tr}$  & 2.5 $\times 10^{-2}$\\
$c$ & 2.67 $\times 10^{-3}$\\
$\alpha$ & -4.37 $\times 10^{-2}$ \\
\hline
\end{tabular}
\end{table}

In principle, the disc shrinkage could be slower, or possibly
delayed if the mass transfer rate from the secondary were reduced,
possibly as a result of a decrease of the illumination of the
secondary. However, when the cooling wave starts, $\alpha$
decreases from $\alpha_{\rm hot}$ to $\alpha_{\rm low}$ in the
outer disc regions, i.e. by a factor 5. $c$ has to be kept fixed,
since one requires the disc to remain eccentric. Because of the
large change in viscosity when a cooling front starts to
propagate, $\dot{M}_{\rm tr}$ would then have to be reduced by an
extremely large factor; we made several numerical experiments with
various prescriptions for a change in $\dot{M}_{\rm tr}$
(amplitude and possible delay with respect to the formation of a
cooling front), and we were unable to find a case in which the
minimum disc radius was not obtained shortly after the end of a
superoutburst.

\section{Possible solutions}

\subsection{ER UMa supercycles}

The standard TTI model cannot reproduce the very short supercycles
observed in ultra-low mass-ratio systems. Osaki (\cite{o95})
reproduced RZ LMI's supercycle by ending prematurely the
superoutburst assuming $r_{\rm crit0} = 0.42 a$ instead of the
usual value of $0.35 a$ (note that in the TTI framework, one
assumes $r_{\rm crit0} = 0.35 a$ when modelling \object{WZ Sge},
even though its mass ratio is presumably smaller than in most SU
UMa stars). In addition, the presence of superhumps after the end
of the superoutburst implies that the eccentricity stops much
later than the end of a superoutburst. The only ways for the disc
to remain eccentric after the end of a superoutburst are then: (i)
$t_{\rm f}$ is very long, or (ii) the accretion disc is always
eccentric in ER UMa systems.

In the first case, the transition time $t_{\rm f}$ must be much
larger than the one used here. However, as shown in Fig.
\ref{fig3}, the interval between a superoutburst and the next
normal outburst is far too long compared to observations. The
second solution implies that the tidal torque is no longer
responsible for superoutbursts, since $c(r) = c_1$ is a constant.
However, the presence of large outbursts is still possible, as the
(modified) DIM predicts the alternation of narrow and wide
outbursts (see paper I) for large enough mass transfer rates. If
the mass transfer rate is constant, the light curve consists of
one or two small outbursts surrounded by large ones, but small
variations of the mass transfer rate can easily lead to ER UMa
type light curves, provided that these variations show some
regularity. Irradiation of the disc and of the secondary can also
account for the presence of long and short outbursts; Hameury et
al. (\cite{hlw00}) have included these effects in the standard
disc instability model and produced light curves typical of
systems such as \object{RZ LMi}.

We therefore conclude that ER UMa stars should be dwarf novae with
a permanently eccentric accretion disc, thereby accounting for
superhumps, and where the illumination of the disc and the
secondary star plays an important role. We thus predict that
superhumps should exist at all phases of the supercycle of ER UMa
stars; this apparently agrees with observations (Gao et al.
\cite{g99}).

\subsection{\label{WZSge} Echo outbursts in WZ Sge stars}

WZ Sge stars have very long supercycles and superoutbursts. In
addition, no normal outbursts between two consecutive
superoutbursts are observed. Some WZ Sge stars also show echo
outbursts at the end of the superoutburst: several small
outbursts, spaced every tens of days, during which superhumps are
still present.

Smak (\cite{s93}) deduces the mass transfer rate of \object{WZ
Sge} from the luminosity of the hot spot. It is observed to
increase at least by a factor 10 during superoutburst, and it
decreases afterwards, remaining larger than the quiescent value
during several tens of days. Such mass transfer rate fluctuation
could result from irradiation of the secondary star by the white
dwarf. Hameury et al. (\cite{hlw00}) have shown that irradiation
(including disc irradiation) could indeed account for some
peculiarities of WZ Sge stars: for example, they reproduced the
echo outbursts phenomenon without including the tidal instability
in the DIM; however, they did not reproduce a full cycle with long
recurrence times, and the echo outbursts they obtained were
slightly too dim.

The long recurrence times can be due to low alpha value (Smak
\cite{s93}; Osaki \cite{o95b}), possibly due to a decay of the MHD
turbulence that would lead to a time-dependant
$\alpha-$prescription (Osaki et al. \cite{o01}), thereby
explaining the echo outbursts. A low viscosity could result from
the secondary being a brown dwarf (Meyer \& Meyer-Hofmeister
\cite{mm99}), but one would have to explain why the viscosity is
so much lower in these systems as compared to other SU UMa systems
which have comparable or even shorter orbital periods. Another
possibility (Lasota et al. \cite{lhh95}, \cite{lkc99}; Hameury et
al. \cite{hlh97}) is that WZ Sge stars are in a stable low state
between superoutbursts, thus explaining the absence of normal
outbursts. This requires a hole in the central regions of the
disc, as a result of either a moderate magnetic field, or of
evaporation. The superoutbursts would then be triggered by an
externally imposed increase of the mass transfer rate, the long
duration of the outburst and the large mass accreted onto the
white dwarf being due to the irradiation of the secondary. They
did not include the tidal instability and, as in the case of ER
UMa stars, the presence of superhumps and late superhumps is
explained if the accretion disc is always eccentric. If one
combines these results with those of Hameury et al. (\cite{hlw00})
on the echo outburst, one should be able to reproduce a WZ Sge
star light curve with echo outbursts, which would however be
slightly different than the observed ones.

We have used our TTI model including irradiation and the presence
of an inner hole for the accretion disc as determined by equation
(7) of Hameury et al. (\cite{hlw00}). We use the parameters of
\object{EG Cnc}. The mass transfer rate from the secondary is
assumed to be affected by irradiation according to:
\begin{equation}
\dot{M}_{\rm tr} = \max(\dot{M}_2,\gamma <\dot{M}_{\rm acc}>)
\label{eq:ill}
\end{equation}
where $\dot{M}_2$ is the mass transfer rate that would be obtained
in the absence of illumination, $\gamma$ is a constant in the
range [0-1], and $<\dot{M}_{\rm acc}>$ is some average of the mass
accretion onto the white dwarf, defined as:
\begin{equation}
<\dot{M}_{\rm acc}> = \int_{-\infty}^{t_0}\dot{M}_{\rm acc}
e^{-(t_0-t)/\Delta t} dt
\end{equation}
$t_0$ being the actual time. This prescription differs slightly
from that of Hameury et al. (\cite{hlh97}), in that the average of
the mass accretion rate is used instead of its value at time
$t_0$; this accounts for the fact that a fraction of the accretion
luminosity is released at later times when the white dwarf cools.
In the following, we have taken $\Delta t$ = 14 days. We start
with a stable system in the low state with a mass transfer rate of
1.5 $10^{15}$ g.s$^{-1}$, in agreement with Smak (\cite{s93})
estimate for WZ Sge ($2. \; 10^{15}$g s$^{-1}$). At $t=10$ days,
the mass transfer rate $\dot{M}_2$ is increased to $3 \;
10^{16}$g.s$^{-1}$; 20 days later, $\dot{M}_2$ is reduced to its
initial value. Our results are insensitive on how $\dot{M}_2$
returns to its normal value, because illumination dominates in Eq.
(\ref{eq:ill}).

This model predicts that the luminosity should increase just
before the triggering of a superoutburst. The increase need not be
large -- one simply requires the mass transfer rate to exceed the
critical rate for stability; this does therefore not contradict
the observations of Ishioka et al. (\cite{i02}) that no strong
orbital hump due to a high mass transfer rate was detected during
the early phase of WZ Sge 2001 outburst.

\begin{figure}
\resizebox{\hsize}{!}{\includegraphics{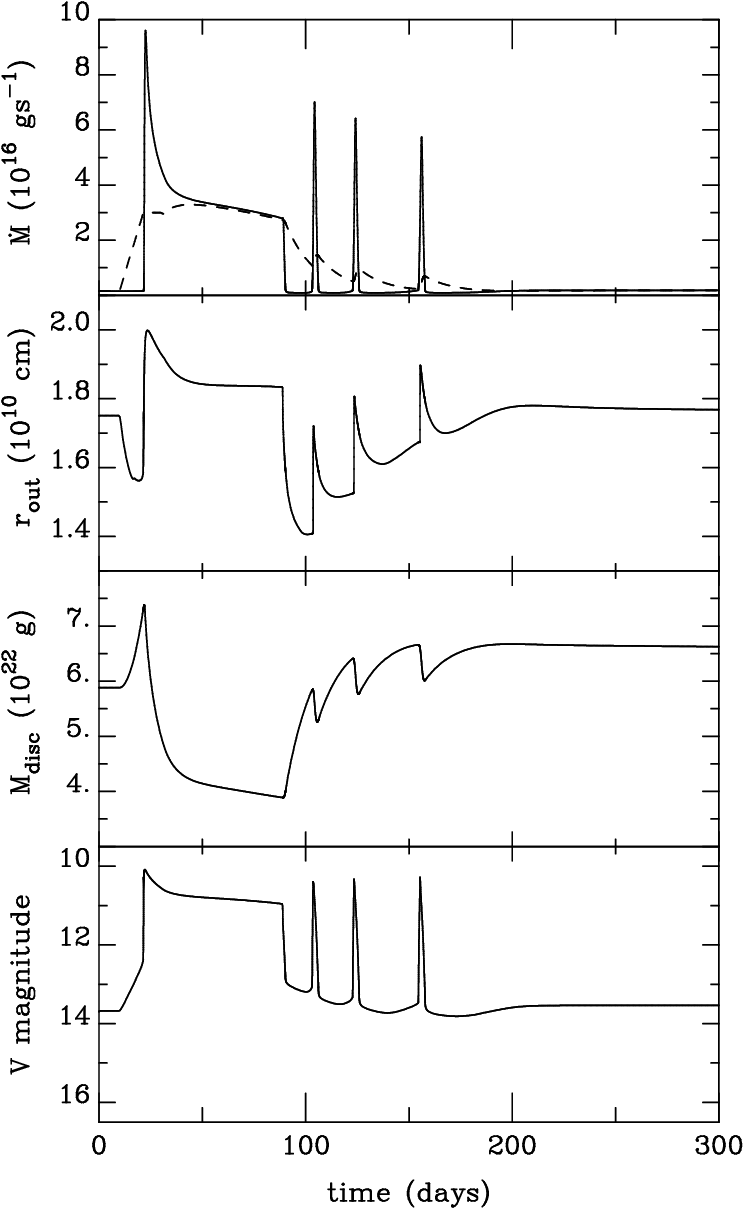}}
\caption{Response of a steady accretion disc to a mass transfer
rate variation by a factor 20. We use the parameters of EG Cnc,
include irradiation, the presence of a hole in the disc, and the
existence of a tidal instability. Initially $\dot{M}_2$ = 1.5
$10^{15}$ g.s$^{-1}$; at $t=10$ $\dot{M}_2$ increases up to 3.0
$10^{16}$ g.s$^{-1}$, and at $t=30$ returns to its quiescent
value. The top panel shows the mass accretion rate onto the white
dwarf (solid line) and the mass transfer rate from the secondary
(dashed line); the other panels show the outer disc radius, its
mass and visual magnitude} \label{fig5}
\end{figure}

We obtain a light curve with a very long superoutburst followed by
3 normal outbursts that last $\sim 3$ days and occur within two
months of the main outburst (Fig. \ref{fig5}). The disc radius
remains always larger than 0.35 $a$, and is always in an eccentric
state; permanent superhumps are thus expected. However, the
interval between normal outbursts is twice longer than for EG Cnc
echo outbursts, which are brighter than observed. We did not
obtain a better fit with \object{ER UMa} light curves; in view of
the very crude assumptions made to derive the mass transfer rate
from the secondary under the influence of illumination, this is
not really a surprise.

\section{Conclusion}

Hellier (\cite{h01}) suggested that several limitations of the TTI
model could be overcome if, in ultra-low mass ratio systems, the
thermal instability ends while the disc is still eccentric. This
would explain the late superhumps, and Hellier (\cite{h01}) also
suggested that this would account for the very short ER UMa
supercycles and the WZ Sge stars echo outbursts.

We do find that the decoupling of the two instabilities can happen
much more easily in ULMR systems than in other SU UMa stars.
However, the tidal instability stops almost immediately after the
end of the thermal-viscous instability (delay less than a day),
because the disc shrinks very rapidly after the departure of a
cooling wave. We estimate that it takes only a few days for the
disc to return to a circular shape after the cessation of the
tidal instability. This is still insufficient to explain late
superhumps that are found tens of days after the superoutburst in
some systems.

We propose instead that the discs in ER UMa systems are
permanently eccentric, and that they should therefore have
permanent superhumps; there are observational indications of this
(Gao et al. \cite{g99}). The short supercycle is then due to
either an alternation of narrow and wide outbursts as in
\object{SS Cyg} (see Buat-M\'enard et al. \cite{b01}), or to the
effects of irradiation (see Hameury et al. \cite{hlw00}). Further
observations of late superhumps should confirm their permanence in
ER UMa stars.

As in Hameury et al. (\cite{hlh97}), we propose that WZ Sge stars
are in a stable cold state during quiescence. The superoutburst is
then triggered by an enhancement of mass transfer rate, and echo
outbursts are triggered while the mass transfer rate is still high
after the superoutburst. However, simulated light curves are still
somewhat different from observations; this might be due to our
crude treatment of the secondary illumination.

\begin{acknowledgements}
This work was supported in part by a grant from {\sl Programme
National de Physique Stellaire} of the CNRS. We thank J.P. Lasota
for a very careful reading of the manuscript and for helpful
comments. We are grateful to the referee, Prof. Y. Osaki, for
pointing us an error in one of the sets of parameters used.
\end{acknowledgements}

\end{document}